\documentclass[aps,prstper,reprint,groupedaddress]{revtex4-2}

\usepackage{graphicx}
\usepackage{dcolumn}
\usepackage{bm}
\usepackage{hyperref}

\begin{document}


\title{Graphical Direct-Writing of Macroscale Domain Structures with Nanoscale Spatial Resolution in Non-Polar-Cut Lithium Niobate on Insulators}



\author{Yuezhao Qian}
\author{Ziqing Zhang}
\author{Yuezhou Liu}
\author{Jingjun Xu}
\author{Guoquan Zhang}
\email[]{zhanggq@nankai.edu.cn}
\affiliation{The MOE Key Laboratory of Weak-Light Nonlinear Photonics, School of Physics and TEDA Applied Physics Institute, Nankai University, Tianjin 300457, China}


\date{\today}

\begin{abstract}
We reported on a graphical domain engineering technique with the capability to fabricate macroscale domain structures with nanoscale spatial resolution in non-polar-cut lithium niobate thin film on insulators through the biased probe tip of scanning atomic force microscopy. It was found that the domain writing process is asymmetric with respect to the spontaneous polarization {\bf{\emph{P}}$_s$} even though the tip-induced poling field is  mirror-symmetric. Various domain structures, with a dimension larger than millimeters while consisting of nanoscale domain elements and with arbitrary domain-wall inclination angle with respect to {\bf{\emph{P}}$_s$}, were designed graphically and then written directly into non-polar-cut lithium niobate crystals. As a proof of principle demonstration, periodically poled x-cut lithium niobate thin film on insulators with a period of 600 nm, a depth of 460 nm  and a length of $\sim$ 1 mm was fabricated. This technique could be useful for device applications in integrated optics and opto-electronics
and domain-wall nanoelectronics based on lithium niobate on insulator. 
\end{abstract}


\maketitle

\section{Introduction}
Lithium niobate (LN) crystal is of excellent electro-optic, pyroelectric, piezoelectric, acousto-optic and nonlinear optical properties, especially with the mature of the fabrication technique of lithium niobate on insulators (LNOI)~\cite{RN306,RN1120,RN1418,RN1523,RN1613}, providing a golden material platform for various ferroelectric, opto-electronic and nonlinear optical applications such as ferroelectric and holographic memories~\cite{RN264,RN1615,RN1616}, electro-optic modulator~\cite{RN1046,RN1128,RN1552}, optical waveguide and integrated optics~\cite{RN1126,RN1523}, and nonlinear optical frequency conversion~\cite{RN1129,RN1534,RN1587,RN1588,RN1617}, to just mention a few. The ferroelectric LN crystal is of 180$^{\rm o}$ domain structure, and its ferroelectric domain can be inverted and designed to improve the device performance, for example, the nonlinear optical frequency conversion efficiency can be improved through periodically poled LN  (PPLN) based on quasi-phase-matching technique~\cite{RN1618,RN739}.  Various domain poling techniques were developed, including the traditional electric field poling technique~\cite{RN348,RN1556}, the light-assisted domain inversion technique~\cite{RN158}, the direct laser pulse and electron beam irradiation techniques~\cite{RN1619,RN775,RN1620,RN1597}, and the tip-field-induced domain inversion technique via scanning probe microscope~\cite{RN1595,RN1267,RN1469,RN1560,RN1537,RN1594,RN1543}. Currently, PPLN with a domain period in micrometers can be easily fabricated in both z-cut and non-polar-cut (x-cut or y-cut) LN crystals. Nanoscale domain structures were also reported in z-cut bulk LN or LNOI, while most of them in bulk LN are surface structures with their size limited by the lateral domain expansion effect during the domain growth~\cite{RN1469,RN1529}. Therefore, up to now, it is still a challenging task to produce PPLN with a sub-micron period in non-polar-cut LN crystals, which is usually required to achieve the parametric interaction of counter-propagating light beams  based on quasi-phase-matching technique and to use the largest nonlinear coefficient $d_{33}$ of LN crystal with improved nonlinear optical frequency conversion efficiency~\cite{RN1591,RN1593}. Tremendous efforts have been put and several groups have tried to fabricate PPLN with sub-micron periodicity in non-polar-cut LN by employing the in-plane electric field poling technique with a finger poling electrode configuration, however, the quality of the sub-micron PPLN still cannot meet the requirement of practical applications such as nonlinear frequency conversion~\cite{RN1542,RN1555,RN1431,RN1587,RN1581}.

More recently, the domain wall of LN was found to be conductive, which  makes it also possible for applications in nanoelectronics~\cite{RN1563,RN469}. The domain-wall conductivity in LN is proportional to $2P_s\sin\theta$, where {\bf{\emph{P}}$_s$} is the spontaneous polarization of LN and $\theta$ is the inclination angle of the domain wall with respect to {\bf{\emph{P}}$_s$}~\cite{RN362}. Unfortunately, the domain-wall inclination angle $\theta$ was found to be almost impossible to control precisely and was small and less than $\rm \sim5^o$ in z-cut LN crystals, which puts a serious limitation on the optimization of domain-wall conductivity in LN~\cite{RN933,RN1458,RN777,RN1502}. Therefore, it is desired to develop a method to fabricate domain walls with large inclination angle $\theta$ in a precisely controllable way.

In this paper, we reported on the successful fabrication of high quality PPLN with a sub-micron periodicity and a longitudinal length longer than millimeters in non-polar-cut LNOI by scanning the biased probe tip of atomic force microscope (AFM). Domain structures with arbitrary shape and domain-wall inclination angle, and with the size ranging from tens of nanometers to millimeters or longer, are predesignable in a graphical way and can then be written directly into non-polar-cut LNOI through the biased AFM tip. This technique is useful in the precise design of domain structures in LNOI for practical applications in nanoelectronics, integrated optics, nonlinear optical frequency conversion and parametric interaction of counter-propagating beams with high efficiency.

\section{Experimental Results}
\begin{figure}[t]
\centering
  \includegraphics[width=8.5cm]{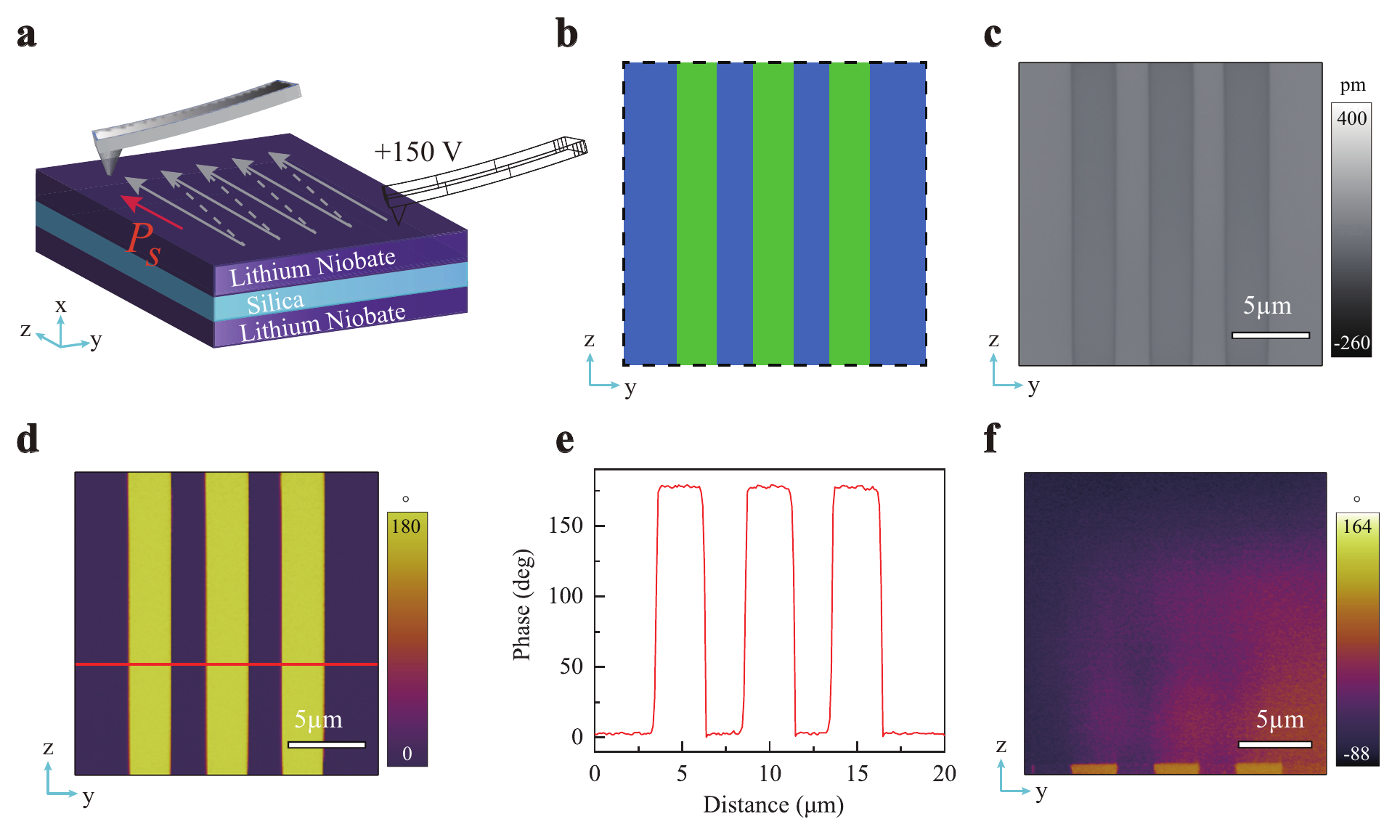}
  \caption{(a) Schematic diagram to fabricate domain structures in x-cut LNOI by scanning the positively biased AFM tip along the spontaneous polarization {\bf{\emph{P}}$_s$}, the grey arrows indicate the scanning direction of the AFM tip. (b) The graphical map used to produce domain stripes, where a positive voltage is applied on the scanning AFM tip in the green regions while no voltage is applied on the AFM tip in the blue regions. (c) The PFM amplitude image of the fabricated domain stripes. (d) The PFM phase image of the fabricated domain stripes. (e) The phase distribution profile of the fabricated domain stripes along the red solid line in (d). (f) The PFM phase image of the x-cut LNOI surface when the positively biased AFM tip scanned anti-parallel to the spontaneous polarization  {\bf{\emph{P}}$_s$} based on the designed voltage map shown in (b). }
  \label{fig:positivevol}
\end{figure}

\begin{figure}[t]
\centering
  \includegraphics[width=8.5cm]{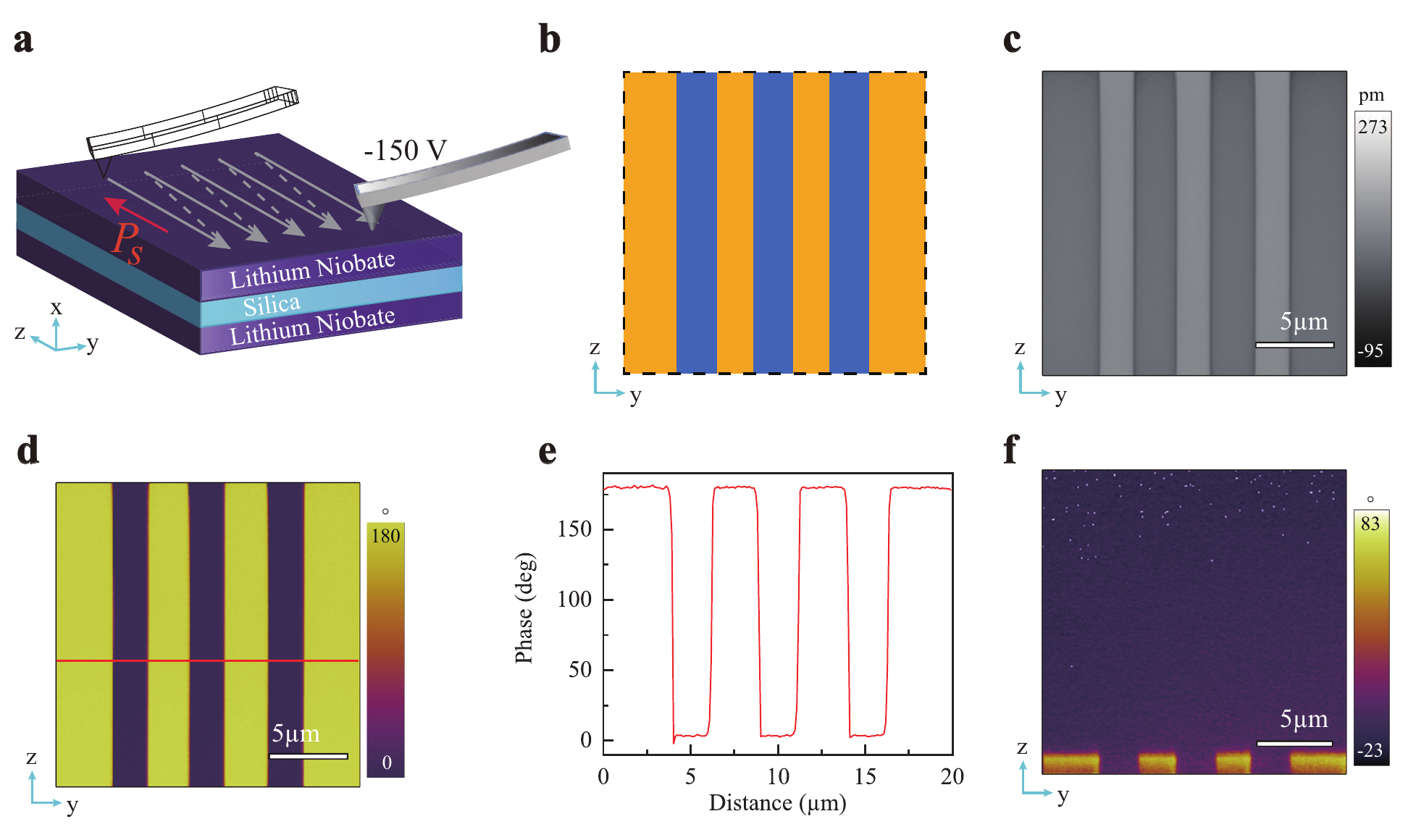}
  \caption{(a) Schematic diagram to fabricate domain structures in x-cut LNOI by scanning the negatively biased AFM tip anti-parallel to the spontaneous polarization {\bf{\emph{P}}$_s$}, the grey arrows indicate the scanning direction of the AFM tip. (b) The graphical map used to produce the domain stripes, where a negative voltage is applied on the scanning AFM tip in the orange regions while no voltage is applied on the AFM tip in the blue regions. (c) The PFM amplitude image of the fabricated domain stripes. (d) The PFM phase image of the fabricated domain stripes. (e) The phase distribution profile of the fabricated domain stripes along the red solid line in (d). (f) The PFM phase image of the x-cut LNOI surface when the negatively biased AFM tip scanned parallel to the spontaneous polarization  {\bf{\emph{P}}$_s$} based on the designed graphical voltage map shown in (b).  }
  \label{fig:negativevol}
\end{figure}

Figure~\ref{fig:positivevol}a shows the schematic diagram to fabricate the domain structures in non-polar-cut LNOI with a positively biased AFM tip. In the experiments, we used a  x-cut LNOI sample consisting of a 600-nm Mg-doped (5.0 mol.\%) x-cut LN thin film, a 2-$\mu$m silica and a 500-$\mu$m LN substrate bonded to each other in sequence. The spontaneous polarization {\bf{\emph{P}}$_s$} of the x-cut LN thin film was in the surface plane of the x-cut LN and was set to be along the z-axis of the experimental coordinate system, as shown in Fig.~\ref{fig:positivevol}a. The AFM tip with a radius of 30 nm was in contact with the top surface of the x-cut LN thin film, and was loaded with a positive voltage and scanned in the direction parallel to the spontaneous polarization {\bf{\emph{P}}$_s$} of the x-cut LN according to a predesigned graphical map, for example, as that shown in Fig.~\ref{fig:positivevol}b, where the AFM tip was positively biased when scanning in the green regions, while the AFM tip was grounded when scanning in the blue regions. Typical domain stripes were fabricated in x-cut LNOI according to the graphical map shown in Fig.~\ref{fig:positivevol}b, in which the AFM tip voltage was 150 V when scanning in the green regions with a scanning speed of 80 $\rm \mu m/s$. The domain stripes were characterized by using piezoresponse force microscopy (PFM, MFP-3D Infinity, Asylum Research, Goleta, CA, USA), and the PFM amplitude and phase images are shown in Figs.~\ref{fig:positivevol}c and~\ref{fig:positivevol}d, respectively. Figure~\ref{fig:positivevol}e is the phase distribution profile along the red solid line in Fig.~\ref{fig:positivevol}d. One sees that the graphical map in Fig.~\ref{fig:positivevol}b is exactly mapped into the domain stripes in x-cut LNOI. However, we found that the domain stripes could not be written into the x-cut LNOI when the AFM tip was scanning in the direction anti-parallel to the spontaneous polarization {\bf{\emph{P}}$_s$}, and only the small area just in front of the scanning tip at the scanning tail terminal was found to be inverted, as shown in Fig.~\ref{fig:positivevol}f.

Surprisingly, the domain stripes can be written into the x-cut LNOI when the AFM tip was scanning in the direction anti-parallel to the spontaneous polarization  {\bf{\emph{P}}$_s$}   but applied with a negative voltage, as shown in Fig.~\ref{fig:negativevol}a. The predesigned graphical map is shown in Fig.~\ref{fig:negativevol}b, in which the AFM tip was applied with a negative voltage when scanning in the orange regions, while it was grounded when scanning in the blue regions. Figures~\ref{fig:negativevol}c and~\ref{fig:negativevol}d show the PFM amplitude and phase images of the fabricated domain stripes when the AFM tip was applied with a negative voltage of $-150$ V and scanning with a speed of 80 $\rm \mu m/s$ according to the graphical map in Fig.~\ref{fig:negativevol}b. The PFM phase distribution profile of the domain stripes along the red solid line in Fig.~\ref{fig:negativevol}d is shown in Fig.~\ref{fig:negativevol}e. Again, one sees that the graphical map in Fig.~\ref{fig:negativevol}b is exactly mapped into the domain stripes in x-cut LNOI. Note that, in contrast to the case with a positively biased AFM tip, in this case the domain stripes could not be written into the x-cut LNOI when the negatively biased AFM tip was scanning in the direction parallel to the spontaneous polarization {\bf{\emph{P}}$_s$}, and only a small part of the domain stripes were written into the x-cut LNOI at the areas just in front of the scanning AFM tip at the scanning tail terminal, as shown in Fig.~\ref{fig:negativevol}f.

\begin{figure}
\centering
  \includegraphics[width=8.5cm]{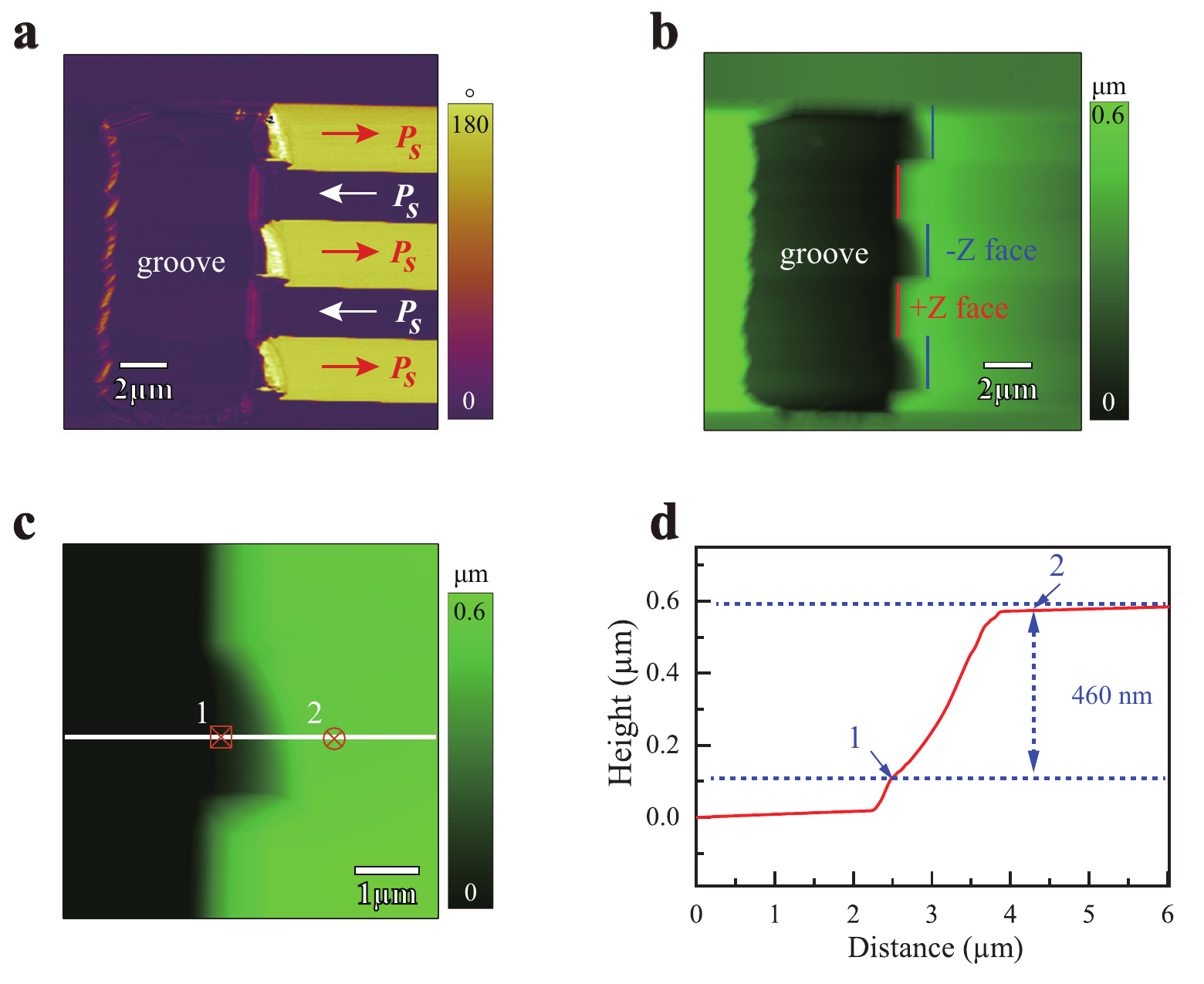}
  \caption{(a) The PFM phase image of the domain stripes fabricated in the x-cut LNOI with a rectangular groove digged by using FIB after a 3-hour HF-acid etching. The red and white arrows represent the polarization direction of the respective domain stripes.  (b) The AFM height map of the grooved domain stripes after a 3-hour HF-acid etching, the red and blue vertical lines indicate the location of the +Z- and -Z-surface of the domain stripes, respectively. (c) The magnified AFM height map of the -Z-surface of the central domain stripe in (b). (d) The height profile across the etched -Z-surface of the domain stripe along the white line in (c), the depth of the inverted domain stripe was measured to be 460 nm.
}
  \label{fig:depth}
 \end{figure}

To measure the depth of the inverted domain stripes, one first fabricated the domain stripes by scanning the negatively biased AFM tip in the direction anti-parallel to the spontaneous polarization {\bf{\emph{P}}$_s$} with a $-150$-V voltage and a scanning speed of 80 $\rm\mu m/s$. Then we digged a rectangular groove with a depth of 600 nm across the domain stripes by employing focused ion beam (FIB) etching, so that the +Z-surface and -Z-surface of the domain stripes were exposed to air. The sample was then immersed in the HF acid with a concentration of 42\% for 3 hours, and the -Z-surface of the domain stripes was etched much faster than the +Z-surface because of the selective etching property of the LN crystals. The PFM phase image and the AFM height map of the etched domain structures are shown in Figs.~\ref{fig:depth}a and~\ref{fig:depth}b. Figure~\ref{fig:depth}c shows the magnified AFM height map around the -Z-surface of the central domain stripe in Fig.~\ref{fig:depth}b, and the height profile of the domain structure along the white line in Fig.~\ref{fig:depth}c,  which is across the etched -Z-surface of the central domain stripe, is shown in Fig.~\ref{fig:depth}d, from which the depth of the inverted domain stripe was measured to be $\sim 460$ nm.

\section{Discussions}
\begin{figure}[h]
 \includegraphics[width=8.5cm]{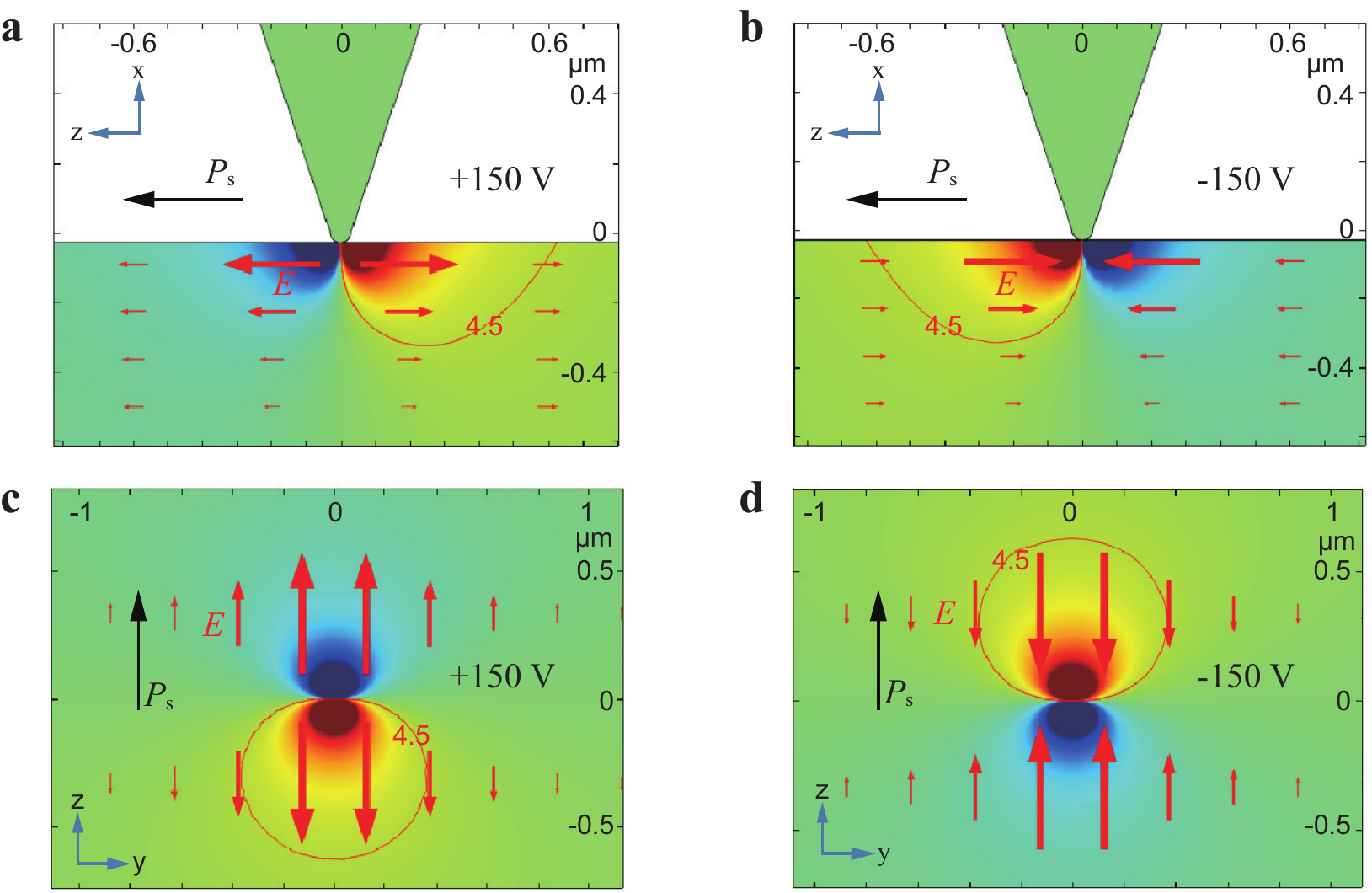}
\caption{ (a) and (c), The spatial distribution of $E_z$ in the x-z and y-z planes, respectively, produced by the positively biased AFM tip with a 150-V voltage. (b) and (d), The spatial distribution of $E_z$ in the x-z and y-z planes, respectively, generated by the negatively biased AFM tip with a $-150$-V voltage. The red arrows represent the tip-induced electric field component $E_z$. The red solid thin curves in all figures are the contour curves of the coercive field with a value of  4.5 kV/mm for Mg-doped LN crystals.}
  \label{fig:field}
\end{figure}

One notes that the domain structures can be directly written into the x-cut LNOI when the positively biased AFM tip scans in the direction parallel to {\bf{\emph{P}}$_s$}, or the negatively biased AFM tip scans in the direction anti-parallel to {\bf{\emph{P}}$_s$}, but not the vice versa. The reason is that the lateral in-plane component of the electric field generated by the biased AFM tip is of opposite direction on the two lobe-sides of the AFM tip. Figure~\ref{fig:field} shows the numerically simulated distribution of the lateral in-plane parallel component of the electric field generated by the biased AFM tip, in which the lateral in-plane parallel electric field component $E_z$ is the electric field component parallel to {\bf{\emph{P}}$_s$} in the x-z or y-z planes. In the simulation, the relative dielectric constant of LN was $\epsilon_{11}=44$ and $\epsilon_{33}=29$, respectively. The radius of the AFM tip was set to be 30 nm, and the tip was in direct contact with the top surface of the x-cut LNOI. The biased voltage was 150 V for Figs.~\ref{fig:field}a and~\ref{fig:field}c, and was $\rm -150$ V for Figs.~\ref{fig:field}b and~\ref{fig:field}d, respectively. The red thin curves in Fig.~\ref{fig:field} show the location where $E_z$ is equal to 4.5 kV/mm, {\it{i.e.}}, the contour curves of the coercive field of Mg-doped LN~\cite{RN348}.  One sees that the electric field component $E_z$ is strong enough to invert the domain polarization around the biased AFM tip, and the depth of the inverted domain can be deeper than 300 nm determined by the contour curves of the coercive field. The experimentally measured depth of the domain stripes is slightly larger than the simulated one determined by the contour curves of the coercive field, this is due to the fact that the domain could grow even when the external electric field is lower than the coercive field, as also observed and confirmed by other groups in LN crystals~\cite{RN1594,RN1543}.

\begin{figure}[t]
  \includegraphics[width=8.5cm]{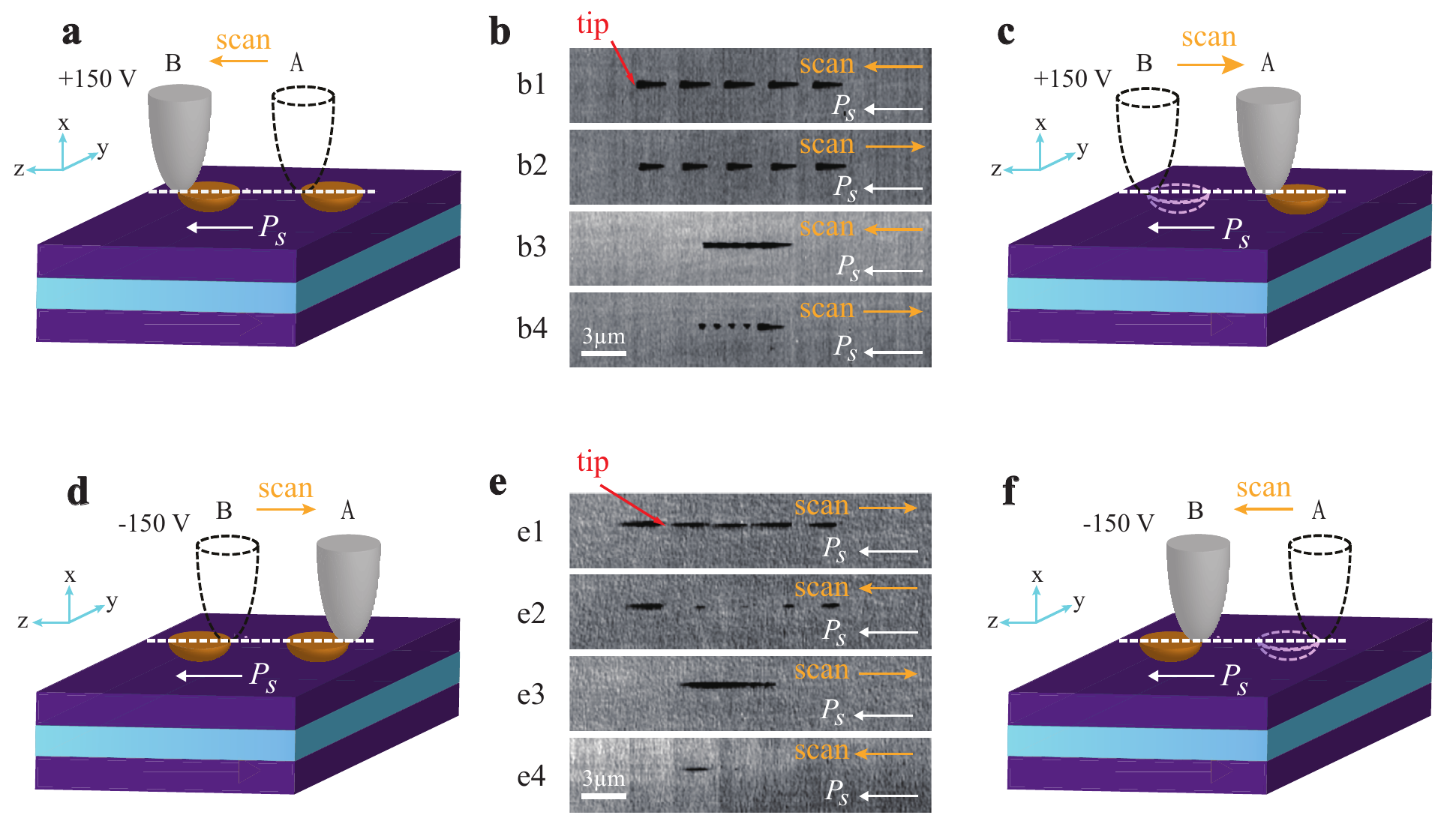}
  \caption{(a) and (c). Schematic diagrams to write the domain structures with a positively biased AFM tip scanning in different directions with respect to  {\bf{\emph{P}}$_s$}. (b) The PFM amplitude images of the domain structures written with a 150-V biased AFM tip, wherein, (b1) and (b3) represent the cases that the tip scanned in the direction parallel to {\bf{\emph{P}}$_s$} but with different neighbouring interval distance of 3 $\mu$m and 1 $\mu$m, respectively. (b2) and (b4) show the cases with the tip scanning in the direction anti-parallel to  {\bf{\emph{P}}$_s$} but with different neighbouring interval distance of 3 $\mu$m and 1 $\mu$m, respectively. (d) and (f). Schematic diagrams to write domain structures with a negatively biased AFM tip scanning in different directions with respect to {\bf{\emph{P}}$_s$}.  (e) The PFM amplitude images of the domain structures generated with a $\rm -150$-V biased AFM tip, wherein, (e1) and (e3) represent the cases that the tip scanned in the direction anti-parallel to {\bf{\emph{P}}$_s$} but with different neighbouring interval distance of 3 $\mu$m and 1 $\mu$m, respectively. (e2) and (e4) show the cases with the tip scanning in the direction parallel to  {\bf{\emph{P}}$_s$} but with different neighbouring interval distance of 3 $\mu$m and 1 $\mu$m, respectively.}
\label{fig:dotdomain}
\end{figure}

The mirror-symmetric distribution of $E_z$ induced by the biased AFM tip would result in the asymmetric behaviour of the domain inversion induced by the biased AFM tip. This is shown in two aspects. First, the domain structure induced by a stationary biased AFM tip is asymmetric. Second, the domain structure formed by the scanning biased AFM tip is asymmetric, and is dependent on the scanning direction with respect to the spontaneous polarization {\bf{\emph{P}}$_s$}. This could be further confirmed experimentally by the following experiments. For example, one wrote a domain structure in the x-cut LNOI with the biased AFM tip at point A, and then lifted the AFM tip and moved the AFM tip to the next point B, there one wrote another domain structure with the biased AFM tip. In this way, one wrote 5 domain structures in sequence with the biased AFM tip with different neighbouring interval distance of 3 $\rm \mu m$ and 1 $\rm \mu m$, respectively, as shown in Fig.~\ref{fig:dotdomain}. One notes that a wedge-like domain structure was formed no matter the AFM tip was positively or negatively biased with a voltage magnitude larger than 65 V, and only the area on one side of the biased AFM tip with its domain polarization anti-parallel to $E_z$ was inverted, while the area on the opposite side of the biased AFM tip with its polarization parallel to $E_z$ was kept to be unchanged, as shown in Figs.~\ref{fig:dotdomain}b1 and~\ref{fig:dotdomain}e1. Similar tip-induced wedge-like domain structures were also observed by other groups in non-polar-cut LN crystals~\cite{RN1594,RN1543}. Interestingly, the wedge-like domain structures are also dependent on the neighbouring interval distance and the scanning direction of the biased AFM tip. For example, the wedge-like domain structures are well isolated from each other when the neighbouring interval distance is large enough, irrespective of the polarity and the scanning direction of the biased AFM tip. On the other hand, when the neighbouring interval distance is small enough, it is possible that the wedge-like domain structure written by the biased AFM tip could be partially or totally erased by the subsequently followed domain writing process, as shown in Figs.~\ref{fig:dotdomain}b2,~\ref{fig:dotdomain}b4,~\ref{fig:dotdomain}e2 and~\ref{fig:dotdomain}e4. This is because the formerly inverted domain polarization was switched back by the electric field induced by the subsequent neighbouring biased AFM tip, for example when the tip was moved from point B to point A in Fig.~\ref{fig:dotdomain}c or from point A to point B in Fig.~\ref{fig:dotdomain}f, due to the fact that the directions of the tip-induced electric field component $E_z$ on the two sides of the biased AFM tip are exactly opposite with respect to each other along the z-axis, as shown in Fig.~\ref{fig:field}. However, the wedge-like domain stripes written by the biased AFM tip in sequence can also be connected one after another and form a long continue domain stripe, as shown in Figs.~\ref{fig:dotdomain}b3 and~\ref{fig:dotdomain}e3, when the positively biased AFM tip scanning in the direction parallel to {\bf{\emph{P}}$_s$} (see Fig.~\ref{fig:dotdomain}a) or the negatively biased AFM tip scanning in the direction anti-parallel to {\bf{\emph{P}}$_s$} (see Fig.~\ref{fig:dotdomain}d) with a small enough neighbouring interval distance. In this way, domain stripes as those shown in Fig.~\ref{fig:positivevol}d and Fig.~\ref{fig:negativevol}d can be directly written via the biased AFM tip.

\begin{figure}[t]
\centering
  \includegraphics[width=8.6cm]{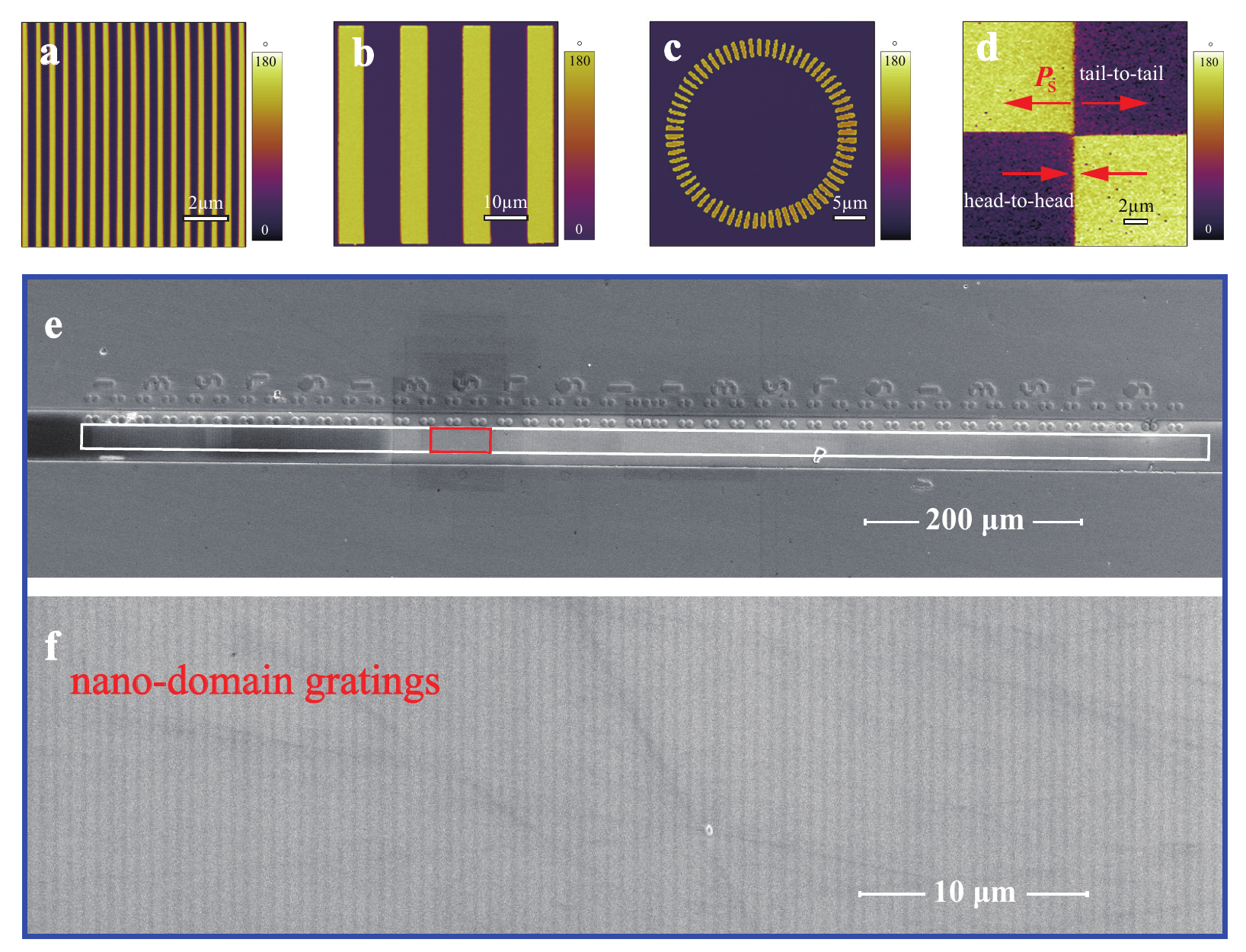}
\caption{Various domain structures fabricated in a 600-nm thick Mg-doped (5.0 mol.\%) x-cut LNOI via a $\rm -150$-V biased AFM tip with a speed of 80 $\rm \mu m/s$ scanning in the direction anti-parallel to the spontaneous polarization {\bf{\emph{P}}$_s$}.  (a) The PFM phase image of PPLN with a grating period of 600 nm. (b) The PFM image of PPLN with a grating period of 15 $\rm \mu m$. (c) The PFM image of domain stripes arranged on a circle with arbitrary domain-wall inclination angle $\theta$. (d) The PFM image of domain structures with head-to-head and tail-to-tail domain walls of a domain-wall inclination angle $\rm \theta= 90^o$. (e) The  scanning electron microscope (SEM) image of PPLN (in the white rectangular box) with a domain period of 600 nm and a longitudinal length of $\sim 1$~mm. (f) The magnified SEM image of nano-domain gratings of PPLN in the red rectangular box in (e).}
\label{fig:arbitrary}
\end{figure}

\begin{figure}[t]
\includegraphics[width=8cm]{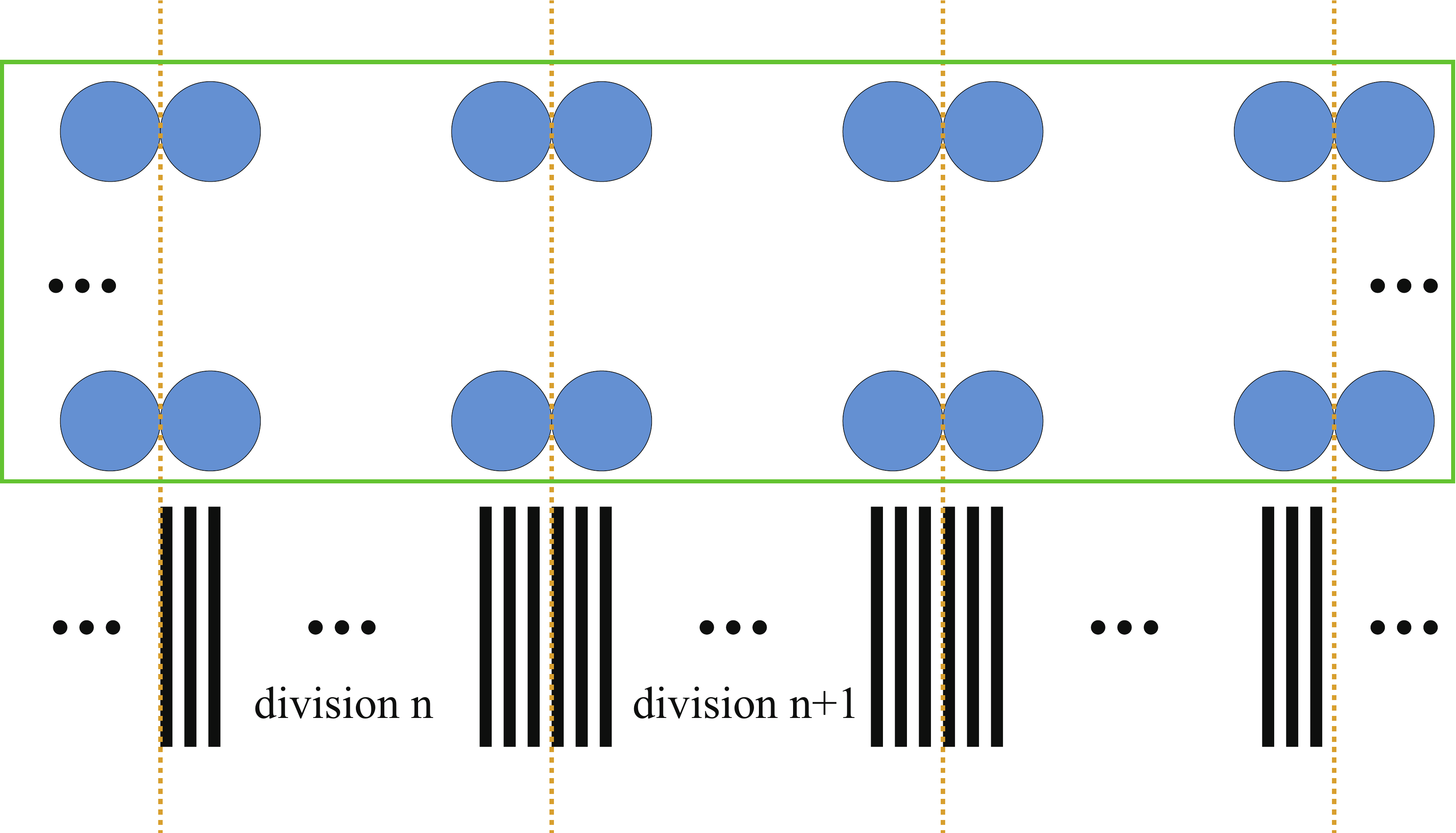}
\caption{The overlay scheme to fabricate macroscale domain structures with a nanoscale spatial resolution via a biased AFM tip of atomic force microscopy.}
\label{fig:overlay}
\end{figure}

With this graphical direct-writing technique, it is capable to design graphically the domain structures and then map onto the non-polar-cut LNOI via the biased AFM tip. Figure~\ref{fig:arbitrary} shows various domain structures fabricated through this graphical direct-writing technique via the biased AFM tip in a 600-nm thick x-cut LNOI, including PPLNs with domain period in sub-micron and micron scale (Figs.~\ref{fig:arbitrary}a and~\ref{fig:arbitrary}b), domain stripes with arbitrary domain-wall inclination angle $\theta$ (Fig.~\ref{fig:arbitrary}c), head-to-head and tail-to-tail domain walls with an inclination angle $\theta=90^o$ (Fig.~\ref{fig:arbitrary}d), and so on. The width of the domain stripes can be down to tens of nanometers ($\sim 70$ nm in our experiments). Figure~\ref{fig:arbitrary}e shows a PPLN with a domain grating period of 600 nm, a domain depth of $\rm \sim$460 nm and a total longitudinal length of $\sim$1 mm.
Note that the scanning area of the AFM probe tip was limited within a scope of $\rm 90\times 90$~$\rm \mu m^2$. To fabricate domain structures long enough in the longitudinal dimension, we employed an overlay technique by dividing the whole area into separate divisions, each division with an area reachable by the AFM tip, and then mapping the domain structure into each division in sequence, as shown in Figs.~\ref{fig:arbitrary}e and~\ref{fig:overlay}. Figure~\ref{fig:arbitrary}f shows the magnified image of the nano-domain gratings in the red box of Fig.~\ref{fig:arbitrary}e, which covers around two divisions, showing the good homogeneity of the fabricated sub-micron PPLN.
Such non-polar-cut sub-micron PPLNs are essential for parametric interaction of counter-propagating light beams and nonlinear conversion generation of ultraviolet light with the feasibility to use the largest nonlinear coefficient $d_{33}$ based on quasi-phase-matching technique. In addition, the capability to fabricate domain walls with arbitrary domain-wall inclination angle in a controllable and designable way also paves the way for applications in domain-wall nanoelectronics based on LNOI~\cite{RN469}.

\section{Conclusion}
In summary, we demonstrated a graphical direct-writing technique of domain structures in non-polar-cut LNOI with a biased AFM tip. We showed the asymmetric behaviour of this domain direct-writing technique, in which the domain can be directly written into the non-polar-cut LNOI when scanning the positively biased AFM tip in the direction parallel to  {\bf{\emph{P}}$_s$} or scanning the negatively biased AFM tip in the direction anti-parallel to {\bf{\emph{P}}$_s$}, but not the vice versa. The tip voltage can be as low as 65 V. The depth of the inverted domain increases with the increase of the biased voltage, and it reaches $\sim$460 nm with a biased voltage of 150 V and a scanning speed of 80 $\rm \mu m/s$. Various domain structures, with the size ranging from nanoscale to millimeter-scale and with precisely controllable domain-wall inclination angle, were designed graphically and then mapped precisely into the non-polar-cut LNOI. As a proof of principle demonstration, PPLN with a domain period of 600 nm, a domain depth of 460 nm and a longitudinal length of $\sim 1$~mm was fabricated in a x-cut, Mg-doped (5.0 mol\%) LNOI, which shows the possibility to use the largest nonlinear coefficient $ d_{33}$ in the nonlinear frequency conversion and parametric interaction of counter-propagating light beams based on quasi-phase-matching technique. In addition, domain structures with head-to-head and tail-to-tail domain walls were also fabricated in the same LNOI sample. The capability to fabricate various domain structures with a nanoscale spatial resolution in a designable and controllable way makes this technique useful for device applications in integrated optics and opto-electronics and domain-wall nanoelectronics.

\section*{acknowledgement}
The authors thank Dr. Xiaojie Wang, Mr. Sanbing Li, Ms Yuchen Zhang and Ms Meili Li from Nankai University for their helpful discussions, and Mr. Wei Wu for the help on the preparation of the FIB-etched samples. This work is supported by the National Key Research and Development Program of China (2019YFA0705000), the National Natural Science Foundation of China (NSFC) (Grant Nos. 12134007, 11774182, 11734009),  and the 111 project (B07013).



%

\end{document}